\documentclass[a4paper]{aa}
\usepackage{graphicx}

\def\degmark{^\circ}
\def \rsun {\ifmmode$R$_{\odot}\else R$_{\odot}$\fi}
\def \nh {N${\rm _H}$}
\def \hcm {\hbox {\ifmmode $ atom cm$^{-2}\else atom cm$^{-2}$\fi}}
\def \src {X\,1658-298}
\def\approxgt{\mathrel{\hbox{\rlap{\lower.55ex \hbox {$\sim$}}
        \kern-.3em \raise.4ex \hbox{$>$}}}}
\def\approxlt{\mathrel{\hbox{\rlap{\lower.55ex \hbox {$\sim$}}
        \kern-.3em \raise.4ex \hbox{$<$}}}}

\newcommand{\SAX}{BeppoSAX}

\newcommand {\Msun} {{ M$_{\odot}$}}

\newcommand {\ergs} {erg~s$^{-1}$}
\newcommand {\ergcms} {erg cm$^{-2}$ s$^{-1}$}
\newcommand {\chisq} {$\chi ^{2}$}
\newcommand {\rchisq} {$\chi_{\nu} ^{2}$}

\begin{document}

\title{BeppoSAX observation of the eclipsing dipping X-ray
binary X\,1658-298}

\author{T. Oosterbroek\inst{1} \and A. N. Parmar\inst{1}  
\and L. Sidoli\inst{1}
\and J. J. M. in 't Zand\inst{2,3}
\and J. Heise\inst{3}
}

\offprints{T. Oosterbroek (toosterb@astro.estec.esa.nl)}

\institute{Astrophysics Division, Space Science Department of ESA, 
ESTEC, P.O. Box 299, 2200 AG Noordwijk, The Netherlands
\and Astronomical Institute, Utrecht University, P.O. Box 80 000, 
3508 TA Utrecht, The Netherlands
\and Space Research Organization Netherlands, Sorbonnelaan 2, 3584 CA Utrecht, The Netherlands
}

\date{Received: 2001 January XX; Accepted: XXXX}

\authorrunning{T. Oosterbroek et al.}

\titlerunning{BeppoSAX observation of \src}

\abstract{ Results of a 2000 August 12--13 BeppoSAX observation of the
7.1~hr eclipsing, dipping, bursting, transient, low-mass X-ray binary (LMXRB)
\src\ are presented. The spectrum outside of eclipses, dips and bursts
can be modeled by the combination of a soft disk-blackbody and a
harder Comptonized component with a small amount $(1.3 \times
10^{21}$~atom~cm$^{-2}$) of low-energy absorption.  In contrast, an
RXTE observation 18 months earlier during the same outburst, measured
an absorption of $5.0 \times 10^{22}$~atom~cm$^{-2}$. Such a change is
consistent with a thinning of the accretion disk as the outburst
progresses.  Structured residuals from the best-fit spectral model are
present which are tentatively identified with Ne-K/Fe-L and Fe-K shell
emission. The spectral changes during dips are complex and
may be modeled by a strong ($\sim$$3 \times 10^{23}$~atom~cm$^{-2}$)
increase in absorption of the Comptonized component only, together with
reductions in normalizations of both spectral components. This
behavior is in contrast to the ``complex continuum''
model for X-ray dip sources, where the softer blackbody
component rapidly suffers strong absorption. It is however, similar to
that found during recent XMM-Newton observations of the eclipsing, dipping,
LMXRB EXO\,0748-676. 
An updated orbital ephemeris is provided which does not
require a quadratic term, in contrast to that of Wachter et 
al. (\cite{w:00}). 
 } \maketitle

\section{Introduction}
\label{sect:introduction}

\src\ is a transient X-ray source discovered by SAS-3 (Lewin et
al. \cite{l:76}). Bursts were reported from \src\ in 1976 October and
1977 June, but there was apparently no significant persistent emission
(Lewin \cite{l:79}). During 1978 March, observations with
SAS-3 and HEAO-1 detected a clearly modulated persistent
emission which showed erratic intensity variations followed by 15
minute duration X-ray eclipses which repeated periodically (Cominsky
\& Wood \cite{cw:84}).  By combining data spanning $\sim$2~years an
orbital period of 7.11~h was obtained (Cominsky \& Wood \cite{cw:89}).
Following the discovery using EXOSAT of a second bursting, dipping, eclipsing,
transient (EXO\thinspace0748-676; Parmar et al. \cite{p:86}) the
region of sky containing \src\ was observed by EXOSAT. However, the
source was not detected. Indeed, until recently \src\ has not been
observed to be active.  In 1991 a 25~ks ROSAT Position
Sensitive Proportional Counter observation failed to detect \src\ with
a 0.2--2.4~keV upper limit of $\sim$$2 \times
10^{-14}$~erg~cm$^{-2}$~s$^{-1}$.  However, on 1999 April~2, after a
21 year quiescent interval, an 18 mCrab (2--9~keV) source at a
position consistent with \src\ was detected in one of the BeppoSAX
Wide Field Cameras (in 't Zand et al. \cite{i:99a}).  
Observations a
day later confirmed that the V=18.3 optical counterpart to \src\
(V2134~Oph; Doxsey et al. \cite{d:79}) had brightened significantly
and was exhibiting high excitation He~{\sc ii} emission, confirming
that the source was undergoing an outburst (Augusteijn et al.\
\cite{az:99}).

\begin{figure*}
\hbox{\hspace{0.0cm}
\includegraphics[height=17.0cm,angle=-90]{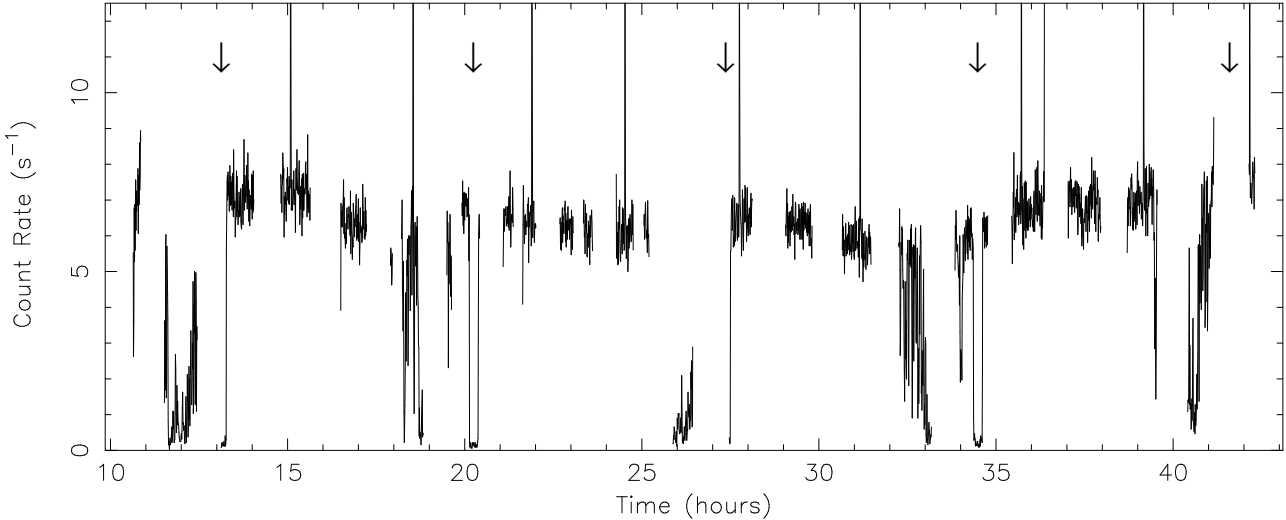}}
  \caption[]{2--10 keV MECS lightcurve of \src\ with a binning of 
32~s. 2 completely and 2 partially observed eclipses, irregular intensity
dips and 9 X-ray bursts are evident. The peak heights of the bursts are
truncated. The arrows mark the expected mid-eclipse times. 
Time is hours of 2000 August 12}
  \label{fig:lc}
\end{figure*}

\src\ was observed by RXTE a few days later (Wachter et
al. \cite{w:00}, hereafter WSB) confirming that the source had resumed
its strong and persistent X-ray emission.  WSB combined their RXTE
eclipse timing measurements with those obtained with HEAO A-1 and SAS-3
and found that the orbital period of \src\ had decreased with a 
timescale of 10$^{7}$~yr. The eclipse transitions of 6 to 13~s were
clearly resolved in the RXTE measurements.  Contemporaneous
observations revealed that the optical modulation is highly variable, 
and exhibits a 0.2 magnitude eclipse feature coincident with the X-ray
eclipse.  Recently, nearly coherent pulsations were discovered
from this source during Type I X-ray bursts (Wijnands et al.
\cite{wij:00}). These imply a spin period of
$\sim$1.8 ms, if the oscillations are directly related to the neutron
star rotation.

Only two Low-Mass X-ray Binary (LMXRB) systems are known that exhibit
both dips and eclipses. (About 10 LMXRBs exhibit only dips).  During
dips the observed spectral changes are complex and inconsistent with
simple absorption from cold material, as might be expected.  Detailed
modeling of these spectral changes provides a powerful
means of studying the structure and location of the emitting and
absorbing regions in LMXRB (e.g., Parmar et al. \cite{p:86}; Church \&
Ba\l uci\'nska-Church \cite{cb:95}; Church et al. \cite{c:97}). This
modeling has revealed that at least two emission components are
required. These regions appear to undergo different amounts of 
absorption during
dips with a point-like blackbody undergoing rapid obscuration
and a cutoff power-law component undergoing a more gradual
increase in obscuration during dipping intervals. 
This gradual change is highly suggestive of an extended component.    
We refer to this as the complex-continuum model.
The same spectral model has been used
to describe the ASCA spectra of a large number of LMXRBs 
(Church \& Ba\l uci\'nska-Church
\cite{cbc:2001}). 

In the case of EXO\thinspace0748-676, (the other eclipsing and
dipping LMXRB), recent XMM-Newton Reflection Grating Spectrometer
(RGS) observations have revealed the presence of narrow spectral
features due to ionized neon, oxygen and nitrogen
(Cottam et al. \cite{c:01}). Simultaneous
European Photon Imaging Camera (EPIC) data revealed that the 0.2--10~keV
spectra can be fit with a two component model consisting of a central
Comptonized component and a more extended thermal halo
(Bonnet-Bidaud et al.\ \cite{bb:01}). In contrast to
the complex continuum continuum model,
the spectral variations observed
during dips can be mainly accounted for by variations in the absorbing
column affecting the Comptonized component only.

\begin{figure*}
\centerline{
\includegraphics[width=8cm,angle=-90]{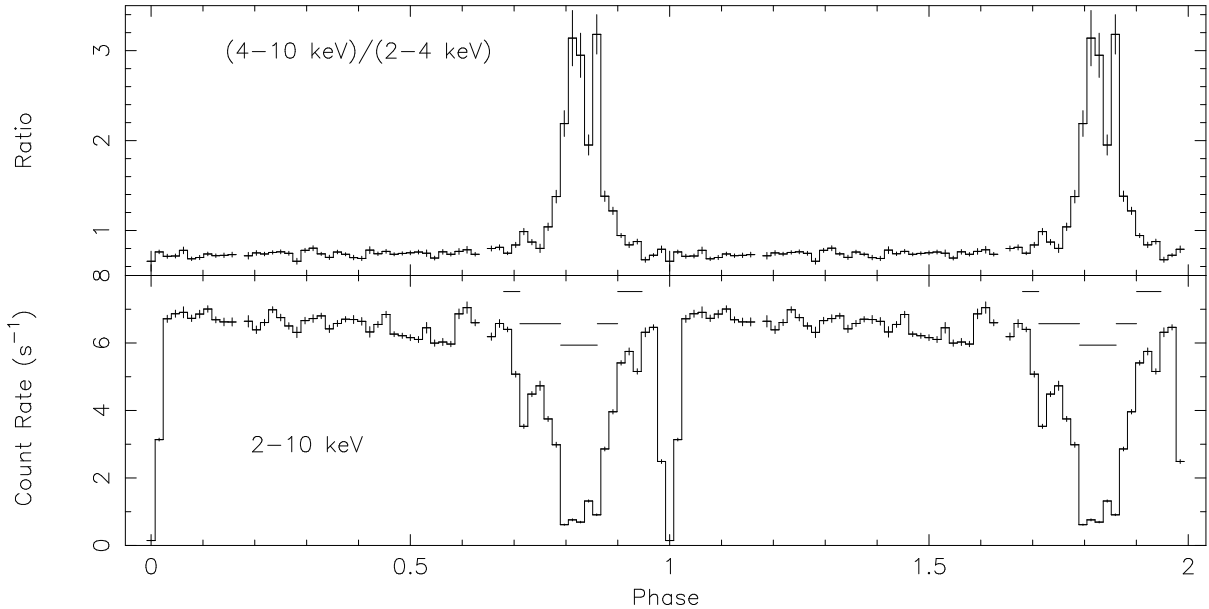}}
\caption[]{The folded hardness ratio (top panel) and
2--10 keV MECS lightcurve showing the deep dipping activity
between phases $\sim$0.68--0.95 and the narrow eclipse centered
on phase 0. X-ray bursts have been removed prior to the folding. 
The horizontal lines indicate phases used to extract the three dip-spectra 
(see Sect.\ \ref{sect:dips}). The central line corresponds
to the deep dip interval,
the two middle lines to the medium dipping, and the two outer lines to the
shallow dipping interval}
\label{fig:ratio}
\end{figure*}

\section{Observations}
\label{sect:observations}

Results from the Low-Energy Concentrator Spectrometer (LECS;
0.1--10~keV; Parmar et al. \cite{p:97}), the Medium-Energy Concentrator
Spectrometer (MECS; 1.8--10~keV; Boella et al. \cite{b:97}),
the High Pressure Gas Scintillation Proportional Counter
(HPGSPC; 5--120~keV; Manzo et al. \cite{m:97}) and the Phoswich
Detection System (PDS; 15--300~keV; Frontera et al. \cite{f:97}) on-board
BeppoSAX
are presented. All these instruments are coaligned and collectively referred
to as the Narrow Field Instruments, or NFI.
The MECS consists of two grazing incidence
telescopes with imaging gas scintillation proportional counters in
their focal planes. The LECS uses an identical concentrator system as
the MECS, but utilizes an ultra-thin entrance window 
to extend the response down to 
0.1~keV. The non-imaging HPGSPC consists of a single unit with a collimator
that was pointed on-source during the whole observation. 
The non-imaging
PDS consists of four independent units arranged in pairs each having a
separate collimator. Each collimator was alternatively
rocked on- and 210\arcmin\ off-source every 96~s during 
the observation.

The region of sky containing \src\ was observed by BeppoSAX on 2000
August 12 10:33 UT to August 13 18:14 UT.  Good data were selected in
the standard way using the SAXDAS 2.0.0 data analysis package.  LECS
and MECS data were extracted centered on the (on-axis) position of
\src\ using radii of 8\arcmin\ and 4\arcmin, respectively.  The
exposure times in the LECS, MECS, HPGSPC, and PDS instruments are 20.2~ks,
53.5~ks, 50.1~ks, and 24.3~ks, respectively.  Background subtraction
for the imaging instruments was performed using standard files, but is
not critical for such a bright source.  Background subtraction for the
HPGSPC was carried out using data obtained when the instrument was
observing the dark Earth and for the PDS using data obtained
when the collimator was offset from the source.

\section{X-ray lightcurve}
\label{subsect:lc}
Fig.\ \ref{fig:lc} shows the 2--10~keV MECS lightcurve of \src\ with a
binning of 32~s spanning just over 4 orbital cycles.  As expected,
eclipses, irregular dips in X-ray intensity, and bursts are
present. Neither the dips nor the eclipses are quite total.  A folded
2--10 keV lightcurve with bursts removed, and the (4.0--10.0
keV/2.0--4.0~keV) hardness ratio as a function of phase are plotted in
Fig.\ \ref{fig:ratio}. From this figure it is evident that strong
dipping behavior occurs between phases 0.68--0.95 (where center of 
X-ray eclipse corresponds to phase 0.0).  
At other phases, the
hardness ratio increases slowly with increasing phase,
from a minimum around phase 0.0.
This may indicate that low-level dipping is present throughout the orbital
cycle. Alternatively, this variation could be due to a dependence of
the X-ray emission on orbital phase.  

\section{Eclipse timing}

The arrival times of the eclipses were determined by fitting a ``step
and ramp'' function (see Parmar et al.\ \cite{p:86}) to the MECS light
curves (which were obtained at 8 s resolution) of the two complete
eclipses. This model has 7 free parameters: the start- and end-times of
the ingress and egress, and the count rates before, during, and after
eclipse. The time of eclipse was calculated as the mean between
the end time of ingress and the start time of egress.

\begin{figure}
\centerline{
\includegraphics[height=7.5cm,angle=-90]{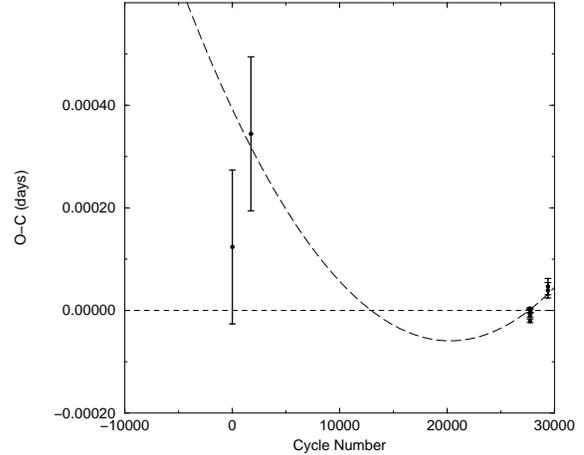}}
\caption[]{The \src\ eclipse timing residuals with respect to 
the best-fit linear ephemeris. The \SAX\
measurements are the two most recent data points.
The best-fit quadratic ephemeris (long dashed line) is also
indicated}
\label{fig:residuals}
\end{figure}

The derived eclipse times, which extend 
the measurements of WSB by more than a year,
or 1600 cycles, are summarized in Table \ref{tab:eclipse_times}.
These times were then combined with the eclipse times compiled by
WSB and linear and quadratic ephemerides investigated.
Fig.~\ref{fig:residuals} shows the residuals with respect to both
ephemerides. Since the WSB results require a quadratic solution
(see below) a quadratic ephemeris was first fit to the eclipse
times to
obtain a $\chi ^2$ of 45.4 for 5 degrees of freedom (dof).  The
updated quadratic ephemeris for the eclipse centers is given by:
\begin{eqnarray}
{\rm T_{ecl}} & = & 2443059.22609(11) + 0.29650453(33)\,{\rm N} \nonumber  \\
             &   & + {\rm (1.1 \pm 0.3)\times 10^{-12} \,N^{2}.} \nonumber
\end{eqnarray}
Note that the sign of the quadratic term is opposite from that of
WSB, and the magnitude a factor $\sim$10 smaller. A
linear ephemeris was next fit to the eclipse times
to give a $\chi ^2$ of 62.9 for 6 dof:
\begin{eqnarray}
{\rm T_{ecl}} & = & 2443059.225826(93) + 0.2965045746(34)\,{\rm N} \nonumber  
\end{eqnarray}
The \chisq\ values of the two fits indicate that the difference 
in \chisq\ is
not significant (the probability of finding a decrease in the \chisq\ by
this value (or larger) is 22\%). Note that the \rchisq\ 
values are much larger than 1, indicating that neither
fit is formally acceptable.

\begin{figure}
\centerline{
\includegraphics[height=7.5cm,angle=-90]{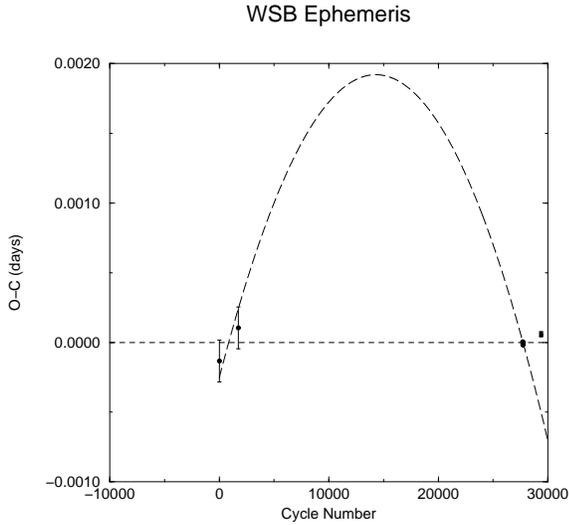}}
\caption[]{The \src\ eclipse timing residuals with respect to a linear
ephemeris obtained using the WSB data {\it only}. The WSB quadratic
ephemeris is indicated by the long-dashed line.
The \SAX\ data points are strongly inconsistent with the WSB quadratic 
ephemeris. Note the different vertical scale with respect to 
Fig.\ \ref{fig:residuals}}
\label{fig:wachter}
\end{figure}

To investigate the large difference in the quadratic terms of the two
ephemerides, {\it only} the data points reported in WSB were
fit. This gives the same solution as WSB found (Fig.\
\ref{fig:wachter}). The WSB solution goes through the two sets of data
points, but deviates strongly from the linear ephemeris in the
interval where no measurements are available. From this we conclude
that the ephemeris obtained by WSB is an artifact of the strongly
non-uniform sampling of the data and that their ${\rm \dot P}$
has no physical relevance.  In Fig.\ \ref{fig:residuals} the
linear ephemeris when the BeppoSAX data are included is plotted. From this
we conclude that a linear ephemeris is an acceptable solution,
although deviations (of the order of a few $\sigma$) are
present. These may be similar to the ``random'' deviations in eclipse
times seen from EXO\thinspace0748-676 (Hertz et al.\ \cite{h:97}).

\begin{table}
\caption[]{New X-ray center of eclipse times (corrected to the solar system
barycenter) for \src. A compilation of previous
measurements can be found in WSB. Cycle
numbers are given using the convention in WSB. 
Uncertainties are given at 68\% confidence}
\begin{tabular}{lll}
\hline
\noalign {\smallskip}
JD$_{\odot}$ & Uncertainty & Cycle \\
\hline
\noalign {\smallskip}
2,451,769.344257  & 0.000016 & 29376 \\
2,451,769.937259  &  0.000015 & 29378 \\
\noalign {\smallskip}                       
\hline
\label{tab:eclipse_times}
\end{tabular}
\end{table}

\section{X-ray spectrum}
\label{subsect:spec}

\subsection{Overall spectrum}

The overall spectrum of \src\ was first investigated by simultaneously
fitting data from all the BeppoSAX NFI. Only intervals when
the source was not bursting, dipping, or eclipsing were included.  
This resulted in
exposure times in the LECS, MECS, HPGSPC, and PDS instruments
of 13.0, 30.8,
27.6, and 13.5~ks, respectively. In order to ensure applicability 
of the \chisq\ statistic, the extracted spectra were
rebinned such that at least 20 counts per bin were present and such
that the resolution of the instrument is not oversampled by more than a
factor 3. LECS data were used in the energy range 0.5--4.0 keV, MECS
data between 1.85--10.5 keV, HPGSPC data between 8--20 keV,
and PDS data in the 15--30 keV energy range, where the instrument responses
are well determined and sufficient counts obtained.  This gives
background-subtracted count rates of 3.5, 6.8, 3.1 and 0.7~s$^{-1}$ in
these energy ranges for the LECS, MECS, HPGSPC, and PDS, respectively.
The photo-electric absorption cross sections of Morrison \& McCammon
(\cite{m:83}) 
were used throughout.  Factors were included in the spectral fitting to
allow for normalization uncertainties between the instruments. These
factors were constrained to be within their usual ranges during the
fitting. All spectral uncertainties and upper-limits are given at 90\%
confidence.  A systematic uncertainty of 1\% was added quadratically
to account for known instrument calibration uncertainties.

Since the source is 7$\degmark$ away from the galactic plane we
verified that emission from the galactic ridge (e.g., Valinia \&
Marshall \cite{vm:98}) is unlikely to affect the background
determination in the non-imaging instruments.  The HPGSPC background
data are obtained during observations of the dark Earth and therefore
the resultant spectrum will be the sum of the source spectrum and the
galactic ridge spectrum. However, given the large offset from the
galactic plane and the brightness of the source, this is unlikely to
be a significant effect. The PDS background spectra are obtained by
rocking the instrument collimator 3\fdg5 in opposite directions.
Thus, changes in the galactic ridge emission (as a function of
galactic latitude) could lead to an imperfect background
subtraction. We therefore accumulated PDS spectra using only the
positive and negative offset background directions. The differences in
the PDS spectra obtained in this way are much smaller than the
statistical uncertainties.

\begin{figure*}
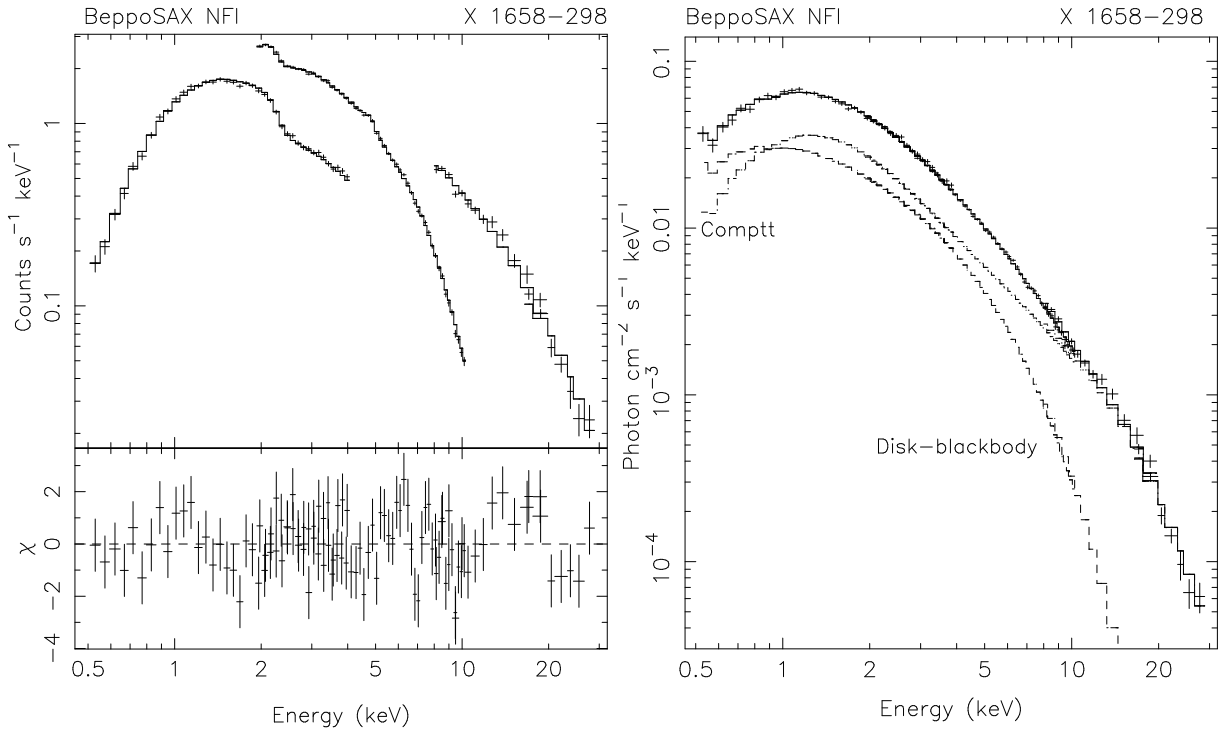

\centerline{\hbox{
\includegraphics[height=8.0cm,angle=-90]{h2628_f5a.ps} 
\includegraphics[height=8.0cm,angle=-90]{h2628_f5b.ps}}}
\caption[]{The 0.5--30~keV
 non-dip, non-burst, and non-eclipse (``persistent emission'')
 \src\ count spectrum and
 residuals when a disk-blackbody and {\sc comptt} model are fit.
 The right hand panel shows the deconvolved photon spectrum with
 the contributions of the two components indicated}
 \label{fig:spectrum}
\end{figure*}

The resulting spectra were first fit with simple models including
absorbed power-laws with photon-index $\alpha$, thermal bremsstrahlung
and cutoff power-laws (${\rm E^{-\alpha}\exp-(E_{c}/kT)}$).  This
showed that no such model provides an acceptable fit with e.g., an
absorbed power-law with $\alpha \sim 2$ giving a $\chi ^2$ of 1149 for
103 dof. A cutoff power-law with $\alpha = 1.3$ and absorption, \nh, of
$3.5 \times 10^{21}$~atom~cm$^{-2}$
provides a significantly better fit with a
$\chi ^2$ of 231 for 102 dof implying the presence of a high-energy
break at 6.1~keV.  
More complex models, consisting of combinations of soft and
hard components were next tried. The soft components tried were blackbodies,
collisionally ionized thermal plasmas (the {\sc mekal} model in {\sc
xspec}), and the disk-blackbody model of Mitsuda et
al. (\cite{m:84}).  
This model assumes that the gravitational energy
released by the accreting material is locally dissipated into
blackbody radiation, that the accretion flow is continuous throughout
the disk, and that the effects of electron scattering are negligible.
There are only two parameters of the model,
$r_{\rm in}({\cos}i)^{0.5}$ where $r_{\rm in}$ is the innermost
radius of the disk, i is the inclination angle of the disk and
$kT_{\rm in}$ the blackbody effective temperature at $r_{\rm in}$.  
The hard components tried were power-laws, cutoff
power-laws and the {\sc xspec} {\sc comptt} Comptonization model
described in Titarchuk (\cite{ti:94}).  This model contains as free
parameters the temperature of the Comptonizing electrons $kT{\rm _e}$,
the plasma optical depth with respect to electron scattering
$\tau_{\rm e}$ and the input temperature of the soft photon (Wien)
distribution $kT_{\rm W}$. A spherical geometry was assumed for the
comptonizing region.  Of all the possible combinations, the best-fit two 
component model is a
disk-blackbody and a {\sc comptt} component with a $\chi ^2$
of 138.6 for 99 dof (Fig.~\ref{fig:spectrum}).  The absorption corrected
2--10~keV luminosity
is $1.6 \times 10^{37}$~erg~s$^{-1}$ for a distance of 15~kpc (Cominsky
\cite{c:81}).
Table~\ref{tab:models}
gives the fit results for models with \rchisq $<$2.
The absorption
derived using the best-fit model of 
$(1.3 \pm 0.2) \times 10^{21}$~atom~cm$^{-2}$ is similar 
to the galactic absorption in the direction of \src\ of
$1.8 \times 10^{21}$~atom~cm$^{-2}$ (Dickey \& Lockman \cite{d:90}).

We note here that we only use the {\sc comptt} model as a
convenient fit model. The physical interpretation of the obtained
parameters (which we do not attempt to do) is difficult. The {\sc
comptt} model assumes the Wien-approximation for the input photons;
for large values of $kT_{\rm 0}$ and low energies this breaks down. In
our case the {\sc comptt} model underestimates the flux up to 40\% at
the lowest energies (see Church \& Ba\l uci\'nska-Church
\cite{cbc:2001}). Additionally, there is evidence that the size of the
Comptonized region is quite large (typically 50000 km; Church
\cite{ch:01}). The temperature of the seed-photons if they originate
from the disk at these distances is much lower than best-fit
seed-temperature for typical luminosities. Additionally, the
difference in shape (at low energies) as discussed above will
influence the normalization and parameters of the second
(disk-blackbody) component, making a physical interpretation of these
parameters even harder.

Interestingly, recent results from the XMM-Newton 
EPIC show that the soft component seen from 
EXO\thinspace0748-676 may be modeled by
a collisionally ionized thermal plasma with strongly
non-solar abundances (Bonnet-Bidaud et al. \cite{bb:01}). This is not
the case in \src, where replacing the disk-blackbody by a
$\sim$1.4~keV {\sc mekal} component with abundances fixed at solar
gives a significantly worse fit with a $\chi ^2$ of 241.8 for 99 dof. Allowing
the abundances to vary does not provide a significantly improved fit.
Additionally, Bonnet-Bidaud et al. (\cite{bb:01}) find that their
EXO\thinspace0748-676 power-law component suffers extra absorption of $\sim$$6
\times 10^{22}$~atom~cm$^{-2}$ during non-dipping or flaring 
intervals.  No such
additional absorption, with a 90\% confidence upper limit of $9 \times
10^{20}$~atom~cm$^{-2}$, is required by the \src\ {\sc comptt}
component during non-dipping intervals.

We next investigated the reasons for the poor ($\chi ^2 = 138.6$ for
99 dof) best-fit.  Inspection of Fig.~\ref{fig:spectrum} reveals the
presence of structured residuals, with amplitudes of $\sim$5\% with
similar forms for the different instruments in the overlapping energy
ranges. This indicates that the observed inadequacies of the spectral
models may be physical, and not caused by incorrectly modeled
instrumental responses. In order to investigate the nature of the
residuals we selected spectra (i) based on count rate, and (ii) based
on orbital phase. The first check was performed in order to see
if the residuals are somehow caused by accumulating data from a 
range of source intensities.
However, no
difference in the structure of the residuals was found between the
high and low count rate selected spectra.  The second test was
performed since it appears that the hardness ratio increases close to
the deep dipping phases (see Fig. \ref{fig:ratio}). Again, no
difference in the structure of the residuals was found between spectra
obtained at ``early'' and ``late'' phases.  We therefore conclude that
the observed residual structure is probably an intrinsic feature of
the \src\ spectrum.

Finally, we tried the ad-hoc addition of Gaussian emission
features. Including narrow features at $1.06 \pm 0.08$ and
$6.23 \pm 0.13$~keV with equivalent widths of $26 \pm 16$ and 
$41 \, \pm \, _{9} ^{17}$~eV reduced the $\chi^2$ to 114.7 for 95 dof. 
The probability of obtaining a decrease in \chisq\ by
this value (or larger) is 1.2\%.
These features may
be tentatively identified with Ne-K/Fe-L and Fe-K shell
emission.  Inspection of the remaining residuals suggests
that other discrete spectral features may be present in the
spectrum. Given that XMM-Newton RGS observations have revealed
O, Ne, and N features in the spectrum of the other eclipsing,
dipping, LMXRB EXO\thinspace0748-676 (Cottam et al. \cite{c:01}), this
is perhaps unsurprising.
We verified that adding similar
Gaussians features to the other 3 models listed in Table~\ref{tab:models}
does not produce markedly better fits. In all cases the change in 
$\chi ^2$ when the Gaussians were included was similar to that of
the best-fit model.

\begin{table*}
\caption[]{\src\ fit results to the non-dip, 
non-eclipse, non-burst spectrum.
\nh\ is in units of $\rm {10^{22}}$ atom $\rm {cm^{-2}}$.
$r{\rm _{in}(\cos i)^{0.5}}$ is in units of km for a distance
of 15~kpc. 90\% confidence limits are given. 
CPL = cutoff power-law, CTT= {\sc comptt}, BB = blackbody, DBB = disk-blackbody. 
All temperatures are in keV. The normalization (``XSPEC units'') of the {\sc comptt}
component in the best-fit (DBB+CTT) 
model is $(3.51\, \pm \, ^{0.30}_{0.48})\times10^{-2}$}
\begin{flushleft}
\begin{tabular}{l@{\hspace{3mm}}c@{\hspace{3mm}}c@{\hspace{3mm}}c@{\hspace{3mm}}c@{\hspace{3mm}}c@{\hspace{3mm}}c@{\hspace{3mm}}c@{\hspace{3mm}}c@{\hspace{3mm}}r}
\hline\noalign{\smallskip}
Model & \hfil N$_{\rm {H}}$ \hfil & $kT_{\rm BB}/kT_{\rm in}$ 
& $r_{\rm in}(\cos i)^{0.5}$ &$\alpha$ & $\rm{E_{c}}$ (keV)  
& $kT_{\rm _W}$ & $kT_{\rm e}$  & ${\rm \tau_e}$ & $\chi^2$/dof \\
\noalign{\smallskip\hrule\smallskip}

BB+CPL& $0.29 \pm 0.03$ & $0.77 \pm 0.07$ &
\dots & $1.25 \pm 0.07 $ & $7.1 \pm 0.9$ & \dots & \dots & 
\dots & 167.1/100\\ 

DBB+CPL& $0.27 \pm 0.03$ & $1.17 \, \pm \, _{0.07} ^{0.15}$ 
 & $ 3.7 \, \pm \, 0.8 $ & $1.09 \pm 0.16$ & 
$7.2 \, \pm \, ^{1.6} _{1.0}$ 
& \dots & \dots & \dots & 158.7/100 \\

BB+CTT & $0.07 \pm 0.02$ & $0.49 \, \pm \, ^{0.01} _{0.04}$ & 
\dots & \dots & \dots & $0.92 \, \pm ^{0.05} _{0.11}$ & 
$4.8 \pm 0.9$ & $3.5 \pm 0.6$ & 144.5/99\\ 

DBB+CTT& $0.13 \pm 0.02$ & $1.63 \pm
0.13$ & $2.3 \, \pm \, ^{0.5} _{0.4}$ & \dots & \dots & $0.39 \pm 0.03 $ 
& $4.0 \, \pm \, ^{0.3} _{0.6}$ & $5.4 \pm 0.6$ & 138.6/99 \\

\noalign{\smallskip\hrule\smallskip}
\end{tabular}
\end{flushleft}
\label{tab:models}
\end{table*}

\subsection{Dip spectrum}
\label{sect:dips}

In order to understand the spectrum of a dip source, a suitable model
must give acceptable fits to both non-dip and dipping intervals. Since
dips are almost certainly due to obscuration by intervening material,
such a model {\it should} fit the dip spectra without requiring any of
the parameters that characterize the shape of the emission, such
as the temperature or power-law slope to be changed. These parameters
are usually obtained by fitting to the non-dip spectrum. In the case
of the ``complex-continuum'' spectral models, which have been used to
model the spectral changes seen in many of the dip sources, the
emission consists of a point-like blackbody plus an extended
Comptonized component. During dips the blackbody is rapidly absorbed,
consistent with it being point-like, while the Comptonized component
is more progressively absorbed, consistent with it being extended
(e.g., Church \& Ba\l uci\'nska-Church \cite{cb:95}; Church et al.
\cite{c:97}). The blackbody is usually assumed to originate from the
surface of the neutron star or from a boundary layer between the
neutron star and an accretion disk, and the extended component from an
isothermal accretion disk corona.

\begin{figure}
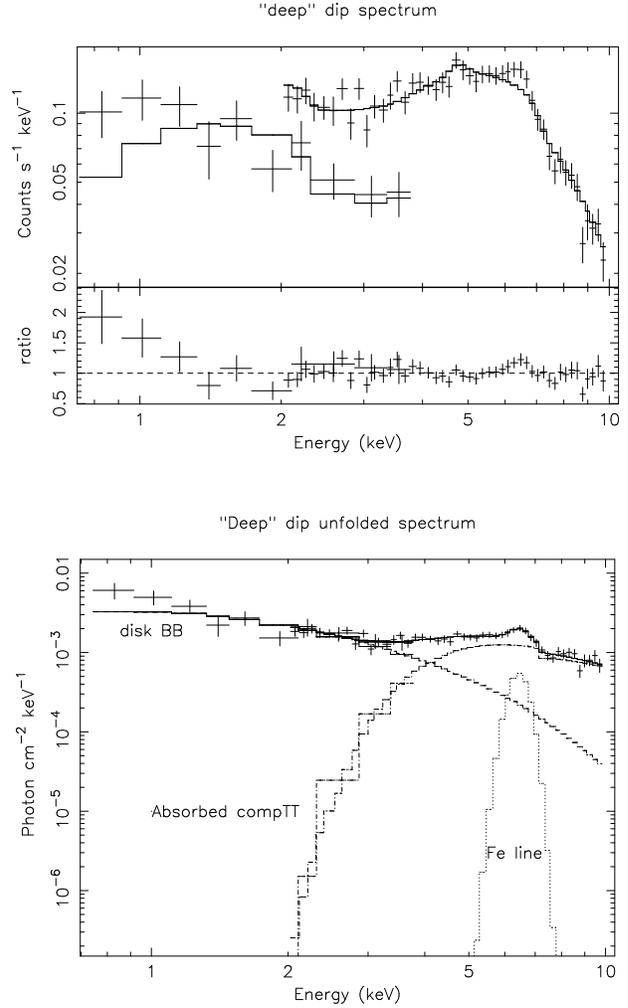

  \centerline{
   	\includegraphics[width=6.0cm,angle=-90]{h2628_f6a.ps}}
	\vspace{.8cm}
   \centerline{
	\includegraphics[width=6.5cm,angle=-90]{h2628_f6b.ps}}
  \caption[]{Top panel: The LECS and MECS spectra of \src\ obtained 
   during the deep dip
   (see text). The fit consists of the model described in the
   text, {\it without} the addition of the Fe-K line to clearly
   show the residuals near 6.4 keV. Lower panel: the photon spectrum
   during the deep dip with the separate components
   indicated. Compare this with Fig.\ \ref{fig:spectrum} to see the large
   effect of absorption on the {\sc comptt} component. The Fe line is
   included in the model for this panel}
  \label{fig:deep_dip_spectrum}
\end{figure}

In order to investigate the spectral changes during \src\ dipping
intervals, three sets of LECS and MECS spectra were generated
corresponding to shallow, medium, and deep dipping intervals (see
Fig.\ \ref{fig:deep_dip_spectrum} for a spectrum during the deep
dip). Selection was made using the orbital phases indicated in Fig.\
\ref{fig:ratio}. The MECS 2--10~keV count rates were 5.4, 3.4, and
0.8~s$^{-1}$ with exposures of 5.2, 5.4, and 4.2~ks, respectively.
Care was taken to minimize the range of count rates included in each
accumulation, since this could lead to ``mixing'' of intervals with
different spectra.  Due to the low count rates and exposure times
HPGSPC and PDS data were not included.

\subsubsection{Simple absorption}
\label{subsubsect:simple}

Initially, the best-fit persistent emission model, modified to allow
additional absorption to both components, was fit to the 
dip spectra. It was not possible to obtain satisfactory fits. 
Inspection of the residuals revealed that significant soft excesses were
present in all 3 dip spectra. 
These can best be modeled if the absorptions 
(${\rm N_H ^{CTT}}$ and ${\rm N_H ^{DBB}}$)
and normalizations  ($I_{\rm CTT}$ and $I_{\rm DBB}$) of the
disk-blackbody and {\sc comptt} components are allowed to vary
separately.
This is the simplest model that is able to adequately model the
spectral changes during dipping intervals and is referred to here as
the ``simple'' absorption model.

\begin{table*}
\caption[]{Results of fitting the simple absorption model 
to the dip spectra. $r_{\rm in}(\cos i)^{0.5}$ assumes
a distance of 15 kpc}
\begin{tabular}{llll}
\hline
\noalign {\smallskip}
Parameter              &  Deep & Medium & Shallow \\
\noalign {\smallskip}
\hline
\noalign {\smallskip}
${\rm N_H ^{CTT}}$ (10$^{22}$ atom cm$^{-2}$) &   $30.8 \pm 4.7$ &
23.1$\pm ^{3.6}_{3.2}$ &
33.6$\pm ^{7.8}_{6.5}$ \\ 
$I_{\rm CTT}$ ($10^{-2}$) & 2.12$\pm^{0.19}_{0.14}$ &  $3.15 \pm 0.23$
& 3.49$\pm ^{0.52}_{0.42}$\\
${\rm N_H ^{DBB}}$ (10$^{22}$ atom cm$^{-2}$) &   $<$0.07 & $<$0.16 & 0.07$\pm0.04$\\
r${\rm _{in}(\cos i})^{0.5}$ & 0.77$\pm0.03$ &2.21$\pm^{0.03}_{0.05}$ & 3.02$\pm$0.03\\
Flux (2--10 keV) (\ergcms)        & $0.95\times10^{-10}$  &
$3.3\times10^{-10}$ & $5.0\times10^{-10}$ \\
$\chi ^2$/dof            & 66.1/52 & 107.2/69 & 74.8/77\\
\noalign {\smallskip}                       
\hline
\label{tab:dip}
\end{tabular}
\end{table*}

\begin{figure}
  \centerline{
   \includegraphics[width=6.0cm,angle=-90]{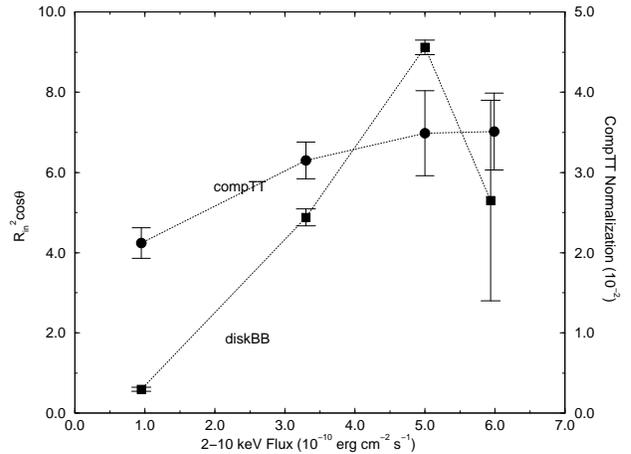}}
  \caption[]{The normalizations of both the disk-blackbody
   (${\rm r^{2}_{\rm in}\cos i}$, left vertical scale, filled squares)
   and {\sc comptt} components (right vertical scale, filled
   circles). From left to right the points correspond to the deep dip,
   medium dip, shallow dip, and persistent spectra.
   The rightmost circle has been shifted by a small
   amount for clarity. See text for a discussion on the value and
   uncertainty for the rightmost (i.e. persistent) disk-blackbody
   normalization. The points are connected with dotted lines to guide
   the eye}
  \label{fig:norms}
\end{figure}

The results of fitting this simple absorption model are reported in 
Table \ref{tab:dip} and show that
during dips the normalizations of both components decrease, while
{\it only} the {\sc comptt} component is affected by strong additional
absorption. 
The depth of dipping is determined primarily by
the reductions in normalization and not by 
an {\it increase} in absorption.
In Fig.\ \ref{fig:norms} the normalizations
of both components are plotted as a function of the total flux in the
dip spectra. This shows that the disk-blackbody normalization has a
stronger dependence on total flux than the
{\sc comptt} component decreasing to a value of $\sim$5\% of the persistent
emission during the deep dip. In contrast, the {\sc comptt} component
has a value $\sim$60\% of the persistent normalization during the deep dip.

We investigated the sudden ``onset'' of the absorption
(i.e.\ even the shallow-dip spectrum shows {\sc comptt} absorption of
$3.4 \times 10^{23}$~atom~cm$^{-2}$)
by fitting the quiescent spectrum with the same simple absorption model.
We were unable to obtain a fit to this spectrum with
strong extra absorption for either component, 
but instead the best-fit was identical to
that reported in Table~\ref{tab:models}. We also experimented with
several other selections to obtain the dip spectra (using only 2 phase
intervals, or using 4 intervals within the dip based on count rate).
All these selections gave similar results to those described above.

We also tried fitting the dip spectra with the same model, but with
either the temperature of the {\sc comptt} component ($kT_{\rm e}$) or
the disk-blackbody temperature ($kT_{\rm in}$) as free parameters. This
gives almost identical results to those given in Table \ref{tab:dip},
with the possible exception of the fit to the deep-dip spectrum when
$kT_{\rm e}$ is treated as a free parameter. This may be applicable if
e.g., different parts of a structured emitting region are covered
during dipping. This gives a \chisq\ of 61.7 for 51 dof with $kT_{\rm
e}$ = 4.6 keV, while the normalization of the {\sc comptt} component
increases to 3.53$\times10^{-2}$, and ${\rm N_H ^{CTT}}$ increases to
$3.6 \times 10^{23}$~atom~cm$^{-2}$.  However, the fact that in only 1
out of 6 trials (three spectra with 2 models) the \chisq\ is reduced
with a significance of 6\% probably means that there is no compelling
reason to treat the temperature of either the {\sc comptt} or
disk-blackbody components as free parameters during dips.

The uncertainty in the value of $r_{\rm in}(\cos i)^{0.5}$ obtained from the fits to
the dip spectra (see Table \ref{tab:dip}) is much smaller than that
obtained from the fits to the persistent emission 
spectrum (Table \ref{tab:models}),
although there are far fewer counts in the dip spectra. This results from
fixing the two temperatures in the model which is used to fit the dip
spectra, while the temperatures are free parameters in the fit to the
persistent emission spectrum. The high value for $r_{\rm in} \cos i$ 
obtained from the
shallow dip spectrum (compared to that obtained from the quiescent
spectrum), is hard to understand physically. However, there is a
strong correlation between $T_{\rm in}$ and $r_{\rm in} \cos i$ and fixing 
$T_{\rm in}$ to 1.76 (best-fit value +90\% confidence uncertainty) 
results in a value for $r_{\rm in} \cos i$ which is within the 
uncertainty quoted in Table~\ref{tab:models}.

We note that the \chisq\ for the medium dip spectrum is
formally unacceptable with a 0.2\% probability that the model
describes the data. 
This might be caused by the
large amount of spectral variability during the intervals 
over which the medium dip spectrum was selected (see Fig.~\ref{fig:lc}). 
This could result in the ``mixing'' of spectra with different
amounts of absorption and normalizations,
which cannot be well fit by the chosen model.

Inspection of the fit residuals for the deep-dip spectrum reveals
clear structure around 6.4 keV, indicative of the presence of an Fe-K
line. The addition of this line reduces the \chisq\ 
from 66.1 for 52 dof to 47.1 for 49 dof, which is significant at $>$99\% 
confidence. The
equivalent width of the line is $330 \, \pm \, ^{200} _{155}$~eV, the
number of photons in the line is $4.4 \times10^{-4}$~cm$^{-2}$~s$^{-1}$, 
the line energy is $6.44 \pm 0.18$~keV, and its width
($\sigma$) is $310 \, \pm \, ^{270} _{310}$~eV.  The only parameter of
the continuum model which changes significantly due to the addition of
the line is the normalization of the {\sc comptt} component, which
decreases by 14\%. 
There is no evidence for lines at 6.44~keV in the
other dip spectra with 1$\sigma$ upper limits to the number of line
photons of 5.0$\times10^{-5}$ cm$^{-2}$ s$^{-1}$ and
7.1$\times10^{-5}$ cm$^{-2}$ s$^{-1}$ for the medium and shallow dip
intervals, respectively. 

\subsubsection{Complex absorption model}

\begin{table}
\caption[]{Fit results of the complex absorption model to the dip
spectra. Units are the same as in Table \ref{tab:dip}}
\begin{tabular}{lllll}
\hline
\noalign {\smallskip}
Parameter              &  Deep & Medium & Shallow \\
\noalign {\smallskip}
\hline
\noalign {\smallskip}
N$_{\rm H}^{\rm DBB}$ & $37.3 \pm 2.5$ & $17.8 \pm 2.3$ & $<$0.09 \\
N$_{\rm H}^{\rm CTT}$  & $123\, \pm \, _{20}^{13}$   & $>$148  
& $13.1 \, \pm \, _{2.2}^{3.4}$ \\
f                     & $0.91 \pm 0.01$        & $0.31 \pm 0.02$   
& $0.47 \pm 0.04$ \\
\chisq /dof           & 70.6/53              & 115.5/70  & 75.7/79   \\
\noalign {\smallskip}                       
\hline
\label{tab:covering}
\end{tabular}
\end{table}

Since fitting the simplest model that provides statistically
acceptable fits to the \src\ dip spectra gives results which are
inconsistent with the physical interpretation of the complex continuum
model of Church \& Ba\l uci\'nska-Church (\cite{cb:95}), where the
softer component is rapidly and strongly absorbed, we also fit this
model to the same dip spectra. Since the blackbody and
cutoff power-law model does not provide a good description of the
persistent emission spectrum (see Table \ref{tab:models}), we instead used
the best-fit {\sc compTT} and disk-blackbody model for this comparison. If the
physical situation is similar, with the disk-blackbody originating from a
compact region and the Comptonized component from an extended region,
this should allow a meaningful comparison to be made with the complex
continuum model.

In this
case the model flux is given by:
\begin{eqnarray}
e^{-\sigma N_H} \;( I_{\rm DBB}  e^{-\sigma N_H^{\rm DBB}}
+\;I_{\rm CTT}\,(f\, e^{-\sigma N_H^{\rm CTT}}+\; (1 - f))), \nonumber
\end{eqnarray}
where $I_{\rm DBB}$ and $I_{\rm CTT}$ are the normalizations of the
disk-blackbody and {\sc comptt} components, $f$ the covering
fraction of the {\sc comptt} component, 
and $\sigma$ the photo-electric absorption cross-sections of Morrison \&
McCammon (\cite{m:83}).
Note that the fundamental differences with the simple absorption model 
that is used in Sect.~\ref{subsubsect:simple} are
that the {\sc comptt} component is partially absorbed and 
$I_{\rm DBB}$ and $I_{\rm CTT}$ are fixed at the values measured
during the persistent emission. 

The best-fit values of \chisq\ are slightly worse than the fit results
from the simple model, but show similar trends, e.g., the medium dip
spectrum is not well modeled (probably because of the same reason as
discussed in Sect.\ \ref{subsubsect:simple}). The absorbing
column for the disk-blackbody component smoothly increases from 
$<$$9 \times 10^{20}$~atom~cm$^{-2}$ during the shallow dip to 
$(3.73 \pm 2.5)\times 10^{23}$~atom~cm$^{-2}$ during 
the deep dip. The trends in the absorption column and
covering fraction of the {\sc comptt} model are not as clear with
the absorbing column increasing from 
$(13.1 \, \pm \, ^{3.4} _{2.2})\times 10^{22}$~atom~cm$^{-2}$ 
during the shallow dip to $>$$10^{23}$~atom~cm$^{-2}$ elsewhere.
The covering fraction ranged from $0.31 \pm 0.02$ during the medium
dip to $0.91 \pm 0.01$ during the deep dip.
The results indicate that at all times the {\sc comptt} component
is obscured by a significantly higher column than the disk-blackbody.
This confirms the same result obtained using the simple absorption model
(see Sect.\ \ref{subsubsect:simple}) and is in contrast to the 
results previously obtained by applying the complex-continuum model
to the dip spectra of other dipping sources.

\subsection{Eclipse spectrum}

\begin{figure}
  \centerline{
   \includegraphics[width=6.0cm,angle=-90]{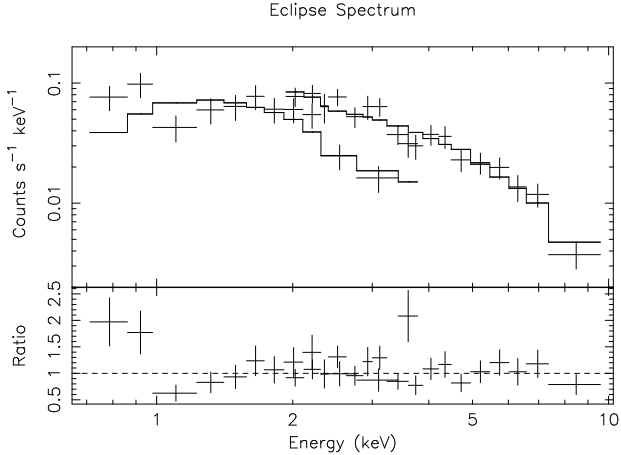}}
  \caption[]{The LECS and MECS eclipse spectrum of \src\ together with
the best-fit absorbed power-law model}
  \label{fig:eclipse_spectrum}
\end{figure}

The spectral shape of the emission during eclipse was investigated. 
Data from all four observed eclipses were combined to
provide exposure times of 1.7 and 2.3~ks for the LECS and MECS,
respectively.  The HPGSPC and PDS spectra do not contain significant
signals and were excluded from this analysis.  The resulting
spectrum can be fit by an absorbed (\nh =
$(3.4 \pm 1.6)\times10^{21}$ \hcm ) power-law with 
$\alpha = 2.0 \, \pm \, ^{0.2}_{0.1}$ (Fig.~\ref{fig:eclipse_spectrum}).
The $\chi ^2$ is 35.3 for 25 dof and the 2--10 keV
luminosity is 4.3$\times10^{35}$~erg~s$^{-1}$ 
for a distance of 15~kpc. 
This is $2.7 \pm 1.0$\% of the persistent value.  Note
the structured residuals in the lowest energy channels of the LECS.
This may suggest the presence of e.g., a dust scattered
component, but the low count rate does not allow us to 
reliably investigate this further.
We note that similar low-energy residual structure may also be present in
the deep dip spectrum (cf. 
Figs.~\ref{fig:deep_dip_spectrum} and \ref{fig:eclipse_spectrum}).
The 90\% confidence upper limit to the number of photons in a 6.44~keV 
emission line is 7.6$\times10^{-5}$~cm$^{-2}$~s$^{-1}$.

\section{X-ray bursts}

Nine X-ray bursts are visible in the light curve (see
Fig.~\ref{fig:lc}). The first burst is the brightest with a peak flux
measured during 2 s intervals of 1.5 times that of the other bursts.
This burst was studied in detail to see if there was evidence for radius
expansion which can be used to estimate the distance to the source.
Light curves were produced in different energy bands. These show that
some structure is present in the burst, especially at higher energies
(which is usually indicative of radius expansion).  Time-sliced
spectra (with 2~s time bins) of the burst where then obtained and the
resulting spectra fit (after subtracting the persistent emission
obtained in a 500~s interval before the burst as background) with a
blackbody model. However, the limited statistics do not allow us to
clearly ascertain whether this burst shows radius expansion. The blackbody
radius during the brightest 2~s of the burst is numerically larger
than the other values, but is consistent (at the 1$\sigma$
level) with being constant. Using a flux estimate obtained from
the two seconds when the burst is brightest (1.4 $\times 10^{-8}$
\ergcms ) and assuming that this corresponds to the theoretical
Eddington limit for a 1.4 \Msun\ neutron star with a hydrogen
atmosphere emitting isotropically a distance of 11~kpc is
obtained.  Using the empirically observed Eddington luminosity
(3.0 $\times 10^{38}$ \ergs) of a sample of bursters for which the
distance is known (Lewin et al.\ \cite{l:93}), a distance of 13~kpc is
obtained.

These estimates agree well with the distance of 15~kpc obtained by
Cominsky (\cite{c:81}) who assumed that the 13 burst sources observed
by SAS-3 were uniformly distributed around the galactic center. 
The distances to individual burst sources, relative to the 
galactic center (assumed to be at 9~kpc), were obtained
by comparing individual properties with those of the sample.
We have searched the literature for reports of other radius expansion
bursts from \src; none were found.
We conclude that the luminosity of the first
burst is likely to be at, or close to, the Eddington limit. Note that
our estimates can be interpreted as upper limits to
the distance, if the burst is not a radius expansion burst (and was
thus emitting at below the Eddington luminosity).

\section{Discussion}
\label{sect:discussion}

We present results of a 2000 August BeppoSAX observation of \src.  The
persistent emission can best be modeled by the summation of
disk-blackbody and Comptonized components with low ($1.3 \times
10^{21}$~atom~cm$^{-2}$) absorption.  The best-fit to a 2--20~keV RXTE
Proportional Counter Array spectrum obtained 16 months earlier is to a
cutoff power-law model with $\alpha = 2.1 \pm 0.1$, \nh$ = (5.0 \pm
0.6) \times 10^{22}$~atom~cm$^{-2}$ and ${\rm E_c} = 8.6 \pm 0.6$~keV
(WSB). When the same model is fit to the BeppoSAX NFI spectrum, the
absorption is only $3.5 \times 10^{21}$~atom~cm$^{-2}$.  A
straightforward interpretation of this striking difference is that
during the early stages of the outburst the X-ray emission is seen
through a strongly thickened accretion disk, while by the time of the
BeppoSAX observation the disk had thinned and the central emission
regions are directly viewed during non-dipping intervals. The observed
2--20 keV luminosity during the RXTE observation of
2.8$\times10^{37}$~erg~s$^{-1}$ (WSB, converted from their quoted flux,
which we assume to be the absorbed value) is indeed $\sim$40\%
higher than our 2--20 keV luminosity, consistent with a higher
mass-accretion rate during the earlier phase of the outburst.

The spectrum of the persistent emission from \src\ can be described by
disk-blackbody and Comptonized components (the {\sc comptt}
model). However the \chisq\ is not formally acceptable and low-level
structure in the residuals is present which is probably due to the
inadequate modeling of the source spectrum. The addition of Gaussian
line features improves the fit (without constraining the parameters
of these features very well). This implies the presence of additional
complexity in the spectrum of \src.  The combination of a
disk-blackbody or blackbody model and a Comptonized component has been
fit a large number of LMXRB spectra (see e.g., Guainazzi et al.\
\cite{g:98}; in 't Zand et al.\ \cite{i:99b}; Barret et al.\
\cite{b:00}; Sidoli et al.\ \cite{s:01}), although a physical
interpretation of the disk-blackbody parameters is not straightforward
(see e.g., Merloni et al.\ \cite{m:00}). It is possible that the
deviations from the smooth continuum in \src\ are caused by
intervening material, through which the emission regions are
observed. The reasons why this intervening material is prominent in
this source is probably due to the high inclination angle - the line
of sight passes close to the surface of the accretion disk. Higher
quality data from {\it Chandra} or XMM-Newton are needed to shed
further light on the nature of these features.

The spectra during dips can be reasonably well approximated with the
persistent model, with only the {\sc comptt} component undergoing
strong additional absorption of $\sim$$3 \times
10^{23}$~atom~cm$^{-2}$. 
The normalizations of both spectral components decreases as the
dipping gets deeper. The normalization of the disk-blackbody component
shows the strongest dependence on total flux, decreasing to $\sim$5\%
of the persistent value during the deep dip, compared to $\sim$60\%
for the {\sc comptt} component. This would suggest
that the disk-blackbody
component originates from a more compact region, while part of the
Comptonized region is still observed directly during dips. 
{\it In this case} it is likely that
during the deepest dipping intervals the disk-blackbody is no longer
viewed directly, the 5\% of emisssion remaining could be X-rays 
scattered into our
line of sight in an extended scattering region (the same region responsible
for the {\sc comptt} component perhaps) in a similar way to the 2.7\%
of emission still visible during eclipse. 
The compact nature of the disk-blackbody and the extended nature
of the Comptonizing region agrees with
results obtained by e.g., Church et al.\ (\cite{c:97}) and Church \&
Ba\l uci\'nska-Church (\cite{cb:95}) with the ``progressive covering''
model. However, the strong additional absorption suffered by the
{\sc comptt} component is in contrast to the results of this
model, where the (disk-)blackbody component 
rapidly undergoes strong additional absorption during dips.

This additional {\sc comptt} absorption is however consistent 
with the picture emerging from XMM-Newton observations of the LMXRB
EXO\thinspace0748-676, where the Comptonized component undergoes
significant extra absorption during dipping intervals (Bonnet-Bidaud
et al.\ \cite{bb:01}), while the softer thermal component does not. However, 
during the EXO\thinspace0748-676 eclipses the thermal component remains,
implying that it is the extended component in this system. 
It is possible to extend this picture to \src\ to obtain a 
consistent picture of {\it all} the spectral changes observed during 
the dips. We first assume that the disk-blackbody component is actually 
an extended `thermal' component similar to that seen by Bonnet-Bidaud
et al.\ (\cite{bb:01}) from EXO\thinspace0748-676 and arranged in a thin
layer above and below the accretion disk. In order that the normalization
of the thermal component reduces to $\sim$5\% of its persistent emission value
during deep dips, $\sim$95\% of the projected emission area should be obscured 
by the dense outer regions of the accretion disk. Assuming that the
disk can be regarded as a dense solid body, then only a small fraction
of the thermal flux will undergo absorption due to skin effects.
This could explain 
why no additional absorption is detected from this component. 
Simulations of the structure of X-ray irradiated disks reveal that 
such sharp transitions in vertical structure may arise due to an ionization
instability (Done \& Nayakshin \cite{dn:01}).

The Comptonized component does, however, suffer from additional absorption
and from a $\sim$40\% reduction in normalization during deep dips. Assuming
that this emission originates from a spherical region centered on the 
neutron star, then during deep dips $\sim$40\% of the emission region 
should be obscured by the solid
accretion disk as with the thermal component. 
The remaining $\sim$60\% is
located away from the line of sight to the accretion disk. It is likely that
the material in the line of sight to the Comptonized component is 
highly turbulent due to the interaction
of the gas stream from the companion star with the outer accretion disk 
and the ionization instability
described by Krolik et al. (\cite{k:81}).
This results in the formation of small
relatively cool clouds embedded in a hotter medium
(Frank et al. \cite{f:87}). The passage of these clouds
in front of the Comptonized region could result
in the observed strong increase in absorption.

While this model is totally ad-hoc, it is an attempt to explain 
the complexity observed during the dips. It does imply a particular
geometry for the emitting regions, and in particular for the thermal
component that must be located close to the accretion disk so that it
can be almost totally obscured. We note that the two systems where this
model may be applicable are the two eclipsing systems which are
seen at inclinations angles of $\sim$75--80$\degmark$, compared to the
pure dip sources which are viewed at inclinations angles of 
$\sim$60--75$\degmark$ (e.g., Frank et al. \cite{f:97}). We speculate
that the different viewing geometry may be one reason for the 
different spectral behavior during dips
of \src\ and EXO\thinspace0748-676 and the other dip sources.

An interesting feature is the appearance of a strong Fe-K
line during the \src\ deep dip, whose energy is consistent with that of
neutral or moderately ionized iron. Its origin (and absence during the other
dip spectra) is rather difficult to understand, but maybe
related to the amount of intervening material in the line of sight where
substantial reprocessing of the continuum radiation can take place.

The BeppoSAX ephemeris does not require a ${\rm \dot P}$ term,
in contrast to the ephemeris of WSB. 
This is probably due to the extremely uneven sampling (and possibly a small
underestimate of the uncertainties) of the arrival times in WSB,
resulting in an artificial solution. We
note that small (random?) deviations from a linear ephemeris are
probably present, but this depends sensitively on the assumed
uncertainties in the arrival times. The observed deviations may be
similar in nature to those observed from EXO\thinspace0748-676 (see Hertz et
al.\ \cite{h:97}).

\acknowledgements 
The BeppoSAX satellite is a joint Italian and Dutch programme. L.
Sidoli acknowledges an ESA Research Fellowship. We thank L. Cominsky
for discussions and the anonymous referee for helpful suggestions.

{}
\end{document}